\documentclass[]{article}  
\usepackage{graphicx,amsmath}
\usepackage{amssymb,amsthm} 
\title{Averaging VMAT treatment plans for multi-criteria navigation}  
\author{David Craft, D{\'a}vid Papp, Jan Unkelbach}        
\date{July 1, 2013}   
\begin{document}
\maketitle

\begin{abstract}
\noindent{\bf Purpose}: To describe a method for combining multiple VMAT plans for use in treatment plan averaging, which is needed for multi-criteria navigation-based treatment planning. \\
{\bf Methods}: We consider sliding window VMAT treatment plans and show that by averaging leaf trajectories in a particular way, the underlying fluence maps are averaged, which in turn yields approximately averaged dose distributions. \\
{\bf Results}: We demonstrate the method on three Pareto optimal plans created for a demanding paraspinal case, where the tumor surrounds the spinal cord. The results show that the leaf averaged plans yield a dose distribution which well-approximates the dosimetric average of the pre-computed Pareto optimal plans. \\
{\bf Conclusions}: This leads to the ability to do direct Pareto surface navigation, i.e. interactive multi-criteria exploration of deliverable VMAT plans.

\end{abstract}

\maketitle

\section{Introduction}

Volumetric modulated arc therapy (VMAT) is a radiation therapy technique which applies continuous irradiation as
the gantry rotates around a patient \cite{yu-vmat-review, otto, gpu-vmat}.  While the gantry is rotating, multi-leaf collimator (MLC) leaves modulate the beam fluence allowing for a highly customized dose distribution. Optimizing these plans has proven challenging \cite{palma}.

Multi-criteria optimization (MCO) via constructing and navigating a Pareto surface has been introduced into the field of radiation therapy to improve the planning time and plan quality for static beam intensity modulated radiation therapy (IMRT) \cite{monz, hoffman, Craft2011}. 
MCO allows planners to interactively explore the dosimetric tradeoffs inherent to radiation therapy planning: more uniform and conformal dose to the target(s) comes at a cost of increased dose to the organs at risk (OARs).  A treatment plan is Pareto optimal if it cannot be improved in any of the planning objectives without worsening some other objective. The Pareto surface navigation method of treatment planning requires a set of Pareto optimal plans to be created, and then during the navigation phase, those plans are averaged to offer a smooth user-driven exploration through the tradeoff space. 

Ideally, as users explore a Pareto surface, they are observing plans that are ready to be delivered. Otherwise, they may navigate to a desired plan, and then in the step that converts that plan into a deliverable version, the plan quality may degrade by an unacceptable amount. For step and shoot IMRT, deliverable navigation can be accomplished by segmenting the pre-computed Pareto database plans and using those segments to form the averaged plans. However, combining pre-segmented plans in this naive way yields plans where the number of segments is equal to the total number of segments from all the contributing plans, which makes such a plan inefficient regarding delivery time. To resolve this issue, multiple approaches to generate segments that are shared across the pre-computed plans have been proposed \cite{delivnav,daomco,fredriksson2013}. 

Research is also taking place to apply Pareto surface navigation to VMAT planning \cite{rasmus-mco-vmat, pardo}. Combining pre-constructed VMAT plans would result in multi-arc plans, one arc for each contributing plan, which would greatly increase the delivery time. The approach in \cite{pardo} gets around this by linearly combining ``modulated blocked arcs'' in order to span a tradeoff space. In \cite{rasmus-mco-vmat}, navigation is initially performed in the fluence map space. After approximating the navigated plan via a VMAT plan, the latter can be fine-tuned by a second navigation in the space of aperture weights.

An alternative approach to combining two VMAT plans consists in averaging their leaf positions at every gantry angle.  This approach will not work for general single arc VMAT plans because of the non-convex (rounded step function) relationship between leaf position and voxel dose. An equivalent way to view this is the following example: if one plan avoided irradiating a critical structure by keeping the leaves on one side of it at a particular gantry angle, and another plan avoided that structure by keeping the leaves on the other side of it, the leaf positions of the combined plan would result in irradiation of the critical structure. 

In this report, we show that under certain restrictions, namely sliding window delivery, VMAT plans can be combined by averaging leaf trajectories in a way different from the one described above, in order to allow deliverable sliding for VMAT MCO.



\section{Methods}
One approach to VMAT planning and delivery utilizes a sliding window technique. In this approach a single $360$ degree arc is divided into $N$ arc sectors. In each arc sector, the leaves move unidirectionally across the field, reversing direction from one sector to the next. Over each arc sector, an effective fluence map is delivered. This is similar to sliding window delivery of IMRT fields, except that for VMAT, the gantry sweeps over the arc sector during the delivery of the fluence map. Multiple approaches to generate sliding window VMAT plans have been proposed \cite{vmerge,networkvmerge,coupled,luan2008,pappvmat}.

A sliding window VMAT plan is characterized by the breakpoints defining the beginning and end of the arc sectors, and the leaf trajectories over all arc sectors. In this section we describe a method to average leaf trajectories of two VMAT plans to create an averaged plan. We show that 
\begin{itemize}
\item[1.] the averaging of leaf trajectories leads to an exact average of the effective fluence map delivered over each arc sector, and 
\item[2.] under realistic assumptions, delivery of the averaged fluence map over the arc sector yields approximately the average dose distribution of the two plans.
\end{itemize}


\subsection{Exact averaging of fluence maps}
\begin{figure}[h!t]
\centering
\includegraphics[trim=0 0 0 0,clip,width=15cm]{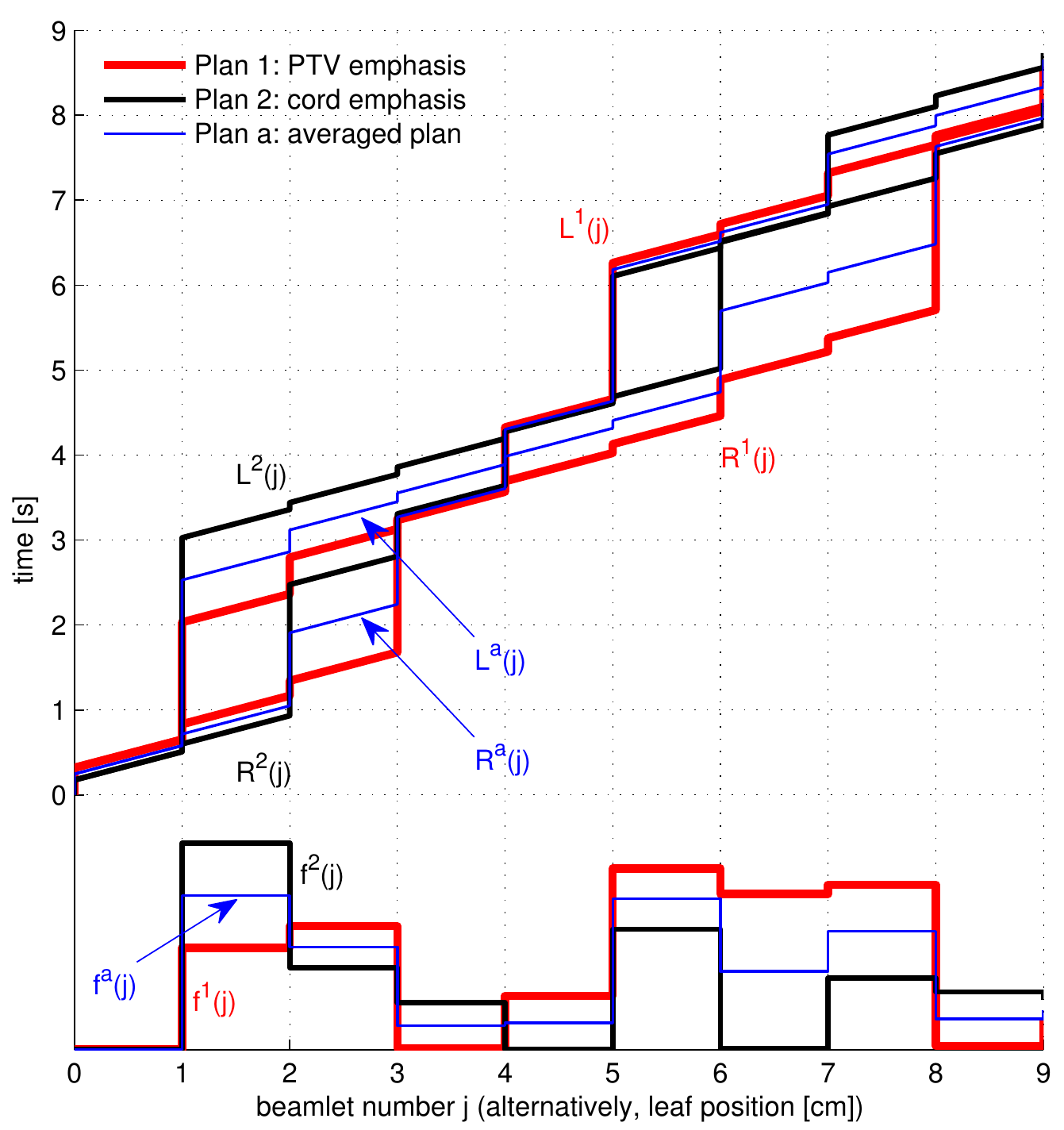}
\caption{Leaf trajectories for a selected arc sector and leaf pair, taken from the paraspinal treatment plan shown in Figure \ref{demo}. The field is divided into 9 beamlets (1 cm wide) and the leaves move across the field in 9 seconds. The red lines correspond to the plan emphasizing PTV coverage, and the black lines displays the plan emphasizing cord sparing. The averaged plan, formed by averaging the left leaf trajectories (along the vertical axis) of the two plans, likewise for the right leaves, is shown in blue. The amount of fluence delivered to a point along the leaf position axis is proportional to the vertical distance between the left and right leaves. Thus, averaged leaf positions yield averaged fluence maps. The effective fluence maps for the three leaf trajectories are shown at the bottom part of the figure.}
\label{sweep}
\end{figure}

For now we will ignore the gantry rotation and focus our attention on a single one of the $N$ fluence map sectors. Focusing further on a single leaf pair, we can supress the leaf index and let the two fluence rows we would like to average be $f_j^1$ and $f_j^2$, where $j$ denotes the beamlet index in leaf travel direction (which can be arbitrarily finely discretized). Each of the fluence maps can be delivered using leaf trajectories $L^1(j)$ and $R^1(j)$ for the leaf and right leaf, respectively. These denote the times in which the left (respectively right) leaf crosses the $j$th leaf position , and thus for a left-to-right sweep, these are strictly increasing functions. This is illustrated in Figure \ref{sweep}.  Note that any optimized fluence map can be converted into leaf trajectories using the standard sliding window leaf sweep algorithm \cite{bortfeld94a}; alternatively, leaf trajectories can directly be optimized \cite{pappvmat}.

Fluence maps are typically expressed in monitor units (MUs), but it is clearer for exposition purposes to assume a fixed (machine maximum) dose rate $\delta$. Thus, time and MU are proportional. Consider a left-to-right sweep. At leaf position $j$ the fluence delivered is proportional to the time duration that $j$ is exposed, which is $L^1(j)-R^1(j)$. Thus we have that $f_j^1 = \delta \left( L^1(j)-R^1(j) \right)$, and $f_j^2 = \delta \left( L^2(j)-R^2(j) \right)$. Therefore, in order to achieve the average fluence map,
we can average the leaf passing times. If we define $L^a(j) = \frac{1}{2}(L^1(j)+ L^2(j))$ and $R^a(j) = \frac{1}{2}(R^1(j)+ R^2(j))$ as the averaged trajectories, we have immediately that 
\begin{align}
f_j^a & = \delta \left( L^a(j)-R^a(j) \right)  \nonumber \\
&= \frac{\delta}{2}(L^1(j)+ L^2(j) - R^1(j) - R^2(j)) \nonumber \\
&= \frac{1}{2}(f_j^1 + f_j^2), \nonumber
\end{align}
i.e. the fluence map is averaged as illustrated in Figure~\ref{sweep}. We note here that if the underlying leaf trajectories are valid from a machine specification point of view (e.g. satisfy the maximum leaf speed constraint), then the averaged trajectories will also be valid.



\subsection{Approximate averaging of dose distributions}
This averaging of leaf trajectories along the time (or MU) axis leads to exact averaging of the fluence profiles, and hence this form of averaging is directly applicable to MCO for sliding window IMRT as well as sliding window VMAT. For VMAT, the caveat is that the fluence maps are delivered over an arc sector, so there is no guarantee that the resulting dose distribution is the average of the underlying dose distributions. The reason for this is that the dose influence matrix (the relationship between beamlet fluence and voxel doses) changes over the arc sector.

We now argue that, under realistic assumptions, delivery of the averaged fluence map will in practice yield approximately the averaged dose distribution of the two original plans. To that end, we note that a given beamlet will be exposed during a short portion of an arc sector. E.g. for a left-to-right sweep, the leftmost beamlet will be exposed in the beginning of the arc sector, whereas the rightmost beamlet will be exposed at the end of the arc sector. As long as for both VMAT plans, all beamlets are exposed during the same portion of the arc sector (i.e. assigned to the same dose-influence matrix) the average fluence map will indeed translate into the average dose distribution. 

In practice, this is approximately fulfilled for the following reason: We are interested in VMAT plans that are deliverable in a short treatment time. Thus the total time is limited, which forces the leaves to traverse the field steadily. In practice, this translates into similar leaf trajectories for different data base plans. This is demonstrated in the Results section.



\section{Results}
\label{sec:results}
To demonstrate the method, we generate a 2D Pareto surface for a paraspinal case, considering the tradeoff between target coverage and spinal cord sparing. We generate three plans on the Pareto surface: the plan that minimizes cord dose, the plan that optimizes target coverage, and a balanced plan. Each VMAT plan is produced using the direct leaf trajectory optimization method described in \cite{pappvmat}. For all plans, the $360$ degree arc is evenly divided into 20 sectors of 18 degrees each. The total delivery time is set to 3 minutes and we assume constant dose rate and gantry speed. Consequently, the leaves have 9 seconds to traverse the field in each arc sector. 

\begin{figure*}[h!t]
\centering
\includegraphics[trim=40 40 40 40,clip,width=\textwidth]{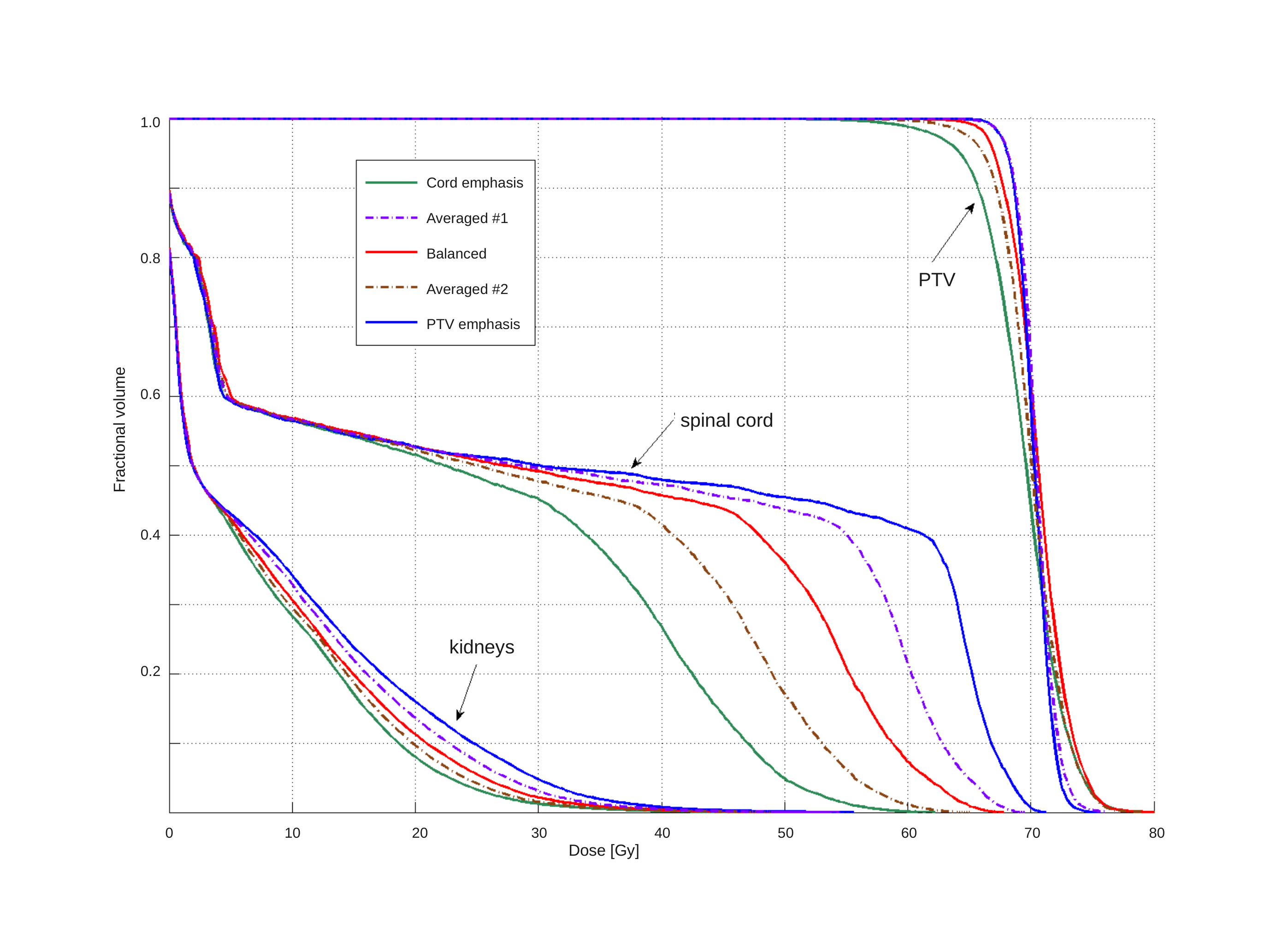}
\caption{Solid DVHs display the three plans that were optimized. The dashed plans are formed by taking leaf trajectory averages as described in the text.}
\label{demo}
\end{figure*}

Figure \ref{demo} displays the dose volume histograms (DVHs) for the three Pareto optimal plans and for two averaged plans. The averaged plans, which are not optimized but computed directly from the trajectory averaging formulas given above, are seen to fall midway between input plans, demonstrating that dosimetric averaging is achieved (we opted to display the compact version of the dose distributions, the DVHs, but we verified that full dose distributions, not shown, also display dosimetric averaging). 

To further illustrate why delivery of the averaged leaf trajectories yields the averaged dose distribution, we consider the leaf trajectories shown in Figure \ref{sweep} in more detail.  For dose calculation and plan optimization, we calculate dose-influence matrices at 2 degree resolution, thus each $18$ degree arc sector is sub-divided into 9 dose calculation sectors. Given 9 second delivery time, each  interval of one second corresponds to one dose calculation sector. In Figure \ref{sweep}, we see that the exposure of one beamlet typically falls onto one or two dose calculation sectors. Furthermore, the beamlet exposure for the two optimized plans and the average plans mostly fall into the same dose calculation sector.

\section{Discussion}
{\it Summary:} Deliverable navigation is a highly sought after property of MCO Pareto surface navigation for radiation therapy planning. Without it, plans can degrade significantly when they are made deliverable, causing an iteration loop of exactly the type that MCO is designed to eliminate. Given that VMAT planning is more challenging than IMRT planning due to the larger dose calculation and the inherent coupling between gantry angles, leaf positions, and dose rates, MCO will be a welcome addition to the field of VMAT planning.  This technical report has presented a method of combining VMAT plans, a key step in Pareto surface navigation based MCO. 

{\it Generalizations:} We have demonstrated averaging two plans, but the method works similarly for a larger number of plans $m$, as happens in higher dimensional Pareto surface navigation. In that case the leaf trajectory formula becomes: $L^a(j) = \sum_{p=1}^m \lambda_p L^p(j)$, where $\lambda_p$ is the weight given to plan $p$. Similarly for $R^a(j)$. 
In the Results section we illustrated the averaging of two VMAT plans that have the same delivery time over each arc sector. However, the presented method generalizes to plans with distinct delivery times. We examined the case of combining two treatment plans optimized under different total treatment time constraints (different by a factor of at most 2), which also yielded dosimetrically averaged plans (not shown).
Furthermore, in the example presented we assumed constant dose rate and constant gantry speed over each arc sector. However, the averaging of leaf trajectories can be generalized by considering cumulative MU (instead of time) as a function of leaf positions.


{\it Delivery efficiency:} Using this averaging method, it is possible that the resulting trajectories are not the time-optimal sliding window deliveries for the yielded fluence map. This is due to the fact that optimal sliding window delivery time is a nonlinear convex function of fluence maps rather than linear. For example, when two fluence maps cancel out each others' peaks and valleys, the optimal sliding window delivery would be faster than the delivery time yielded by the simple method described herein. However, given that plans that are to be combined during Pareto navigation are not wildly different (fluence maps peaks and valleys similar in location and magnitude), this discrepency between linearity and convexity is much reduced. Also, the fact that there might be a faster way to sweep the average fluence map does not diminish the value of this work, which is to provide a practical method for combining VMAT plans during MCO navigation.

{\it Future work:} As this is a brief technical note, we have not examined how the algorithm behaves as the size of the arc sectors increase or as the irradiated volume grows. These cases potentially lead to more variability in the leaf trajectories, and one expects the resulting dose distribution to deviate increasingly from the dosimetrically averaged one. Further work is needed to validate the accuracy of our method for a variety of patient geometries. However, given that high quality VMAT plans using sliding window approaches usually use 18 degrees or less for the size of the fluence sectors, and that for this value our results are promising, even for a highly concave dose distribution of a paraspinal case, we do not expect this to compromise the practical usefulness of the proposed method. 

 
\bigskip
\bibliographystyle{unsrt}

\end{document}